\documentclass[aps,nofootinbib,amsmath,prd,twocolumn,showpacs,superscriptaddress,groupedaddress,floatfix,preprintnumbers]{revtex4}  

\usepackage{graphicx}  
\usepackage{dcolumn}   
\usepackage{bm}        
\usepackage{amssymb}   

\usepackage{graphics,color,array}
\usepackage{xspace}

\newcommand{\pmu}[1]{\partial_{\mu} #1}
\newcommand{\pnu}[1]{\partial_{\nu} #1}
\newcommand{\pMu}[1]{\partial^{\mu} #1}
\newcommand{\pNu}[1]{\partial^{\nu} #1}
\newcommand{\Pmu}[1]{\nabla_{\mu} #1}
\newcommand{\Pnu}[1]{\nabla_{\nu} #1}
\newcommand{\PMu}[1]{\nabla^{\mu} #1}
\newcommand{\PNu}[1]{\nabla^{\nu} #1}

\begin{document}

\hspace{13.5cm} \mbox{MZ-TH/12-28}

\title{Functional renormalization group of the nonlinear sigma model \\ and the ${\rm O}(N)$ universality class.}

\author{Raphael Flore}
\email{raphael.flore@uni-jena.de}
\affiliation{Theoretical-Physical Institute, Friedrich-Schiller-University Jena, Max-Wien-Platz 1, D-07743 Jena, Germany}

\author{Andreas Wipf}
\email{wipf@tpi.uni-jena.de}
\affiliation{Theoretical-Physical Institute, Friedrich-Schiller-University Jena, Max-Wien-Platz 1, D-07743 Jena, Germany}

\author{Omar Zanusso}
\email{zanusso@thep.physik.uni-mainz.de}
\affiliation{Institute of Physics, Johannes-Gutenberg-University Mainz, Staudingerweg 7, D-55099 Mainz, Germany}


\begin{abstract}
We study the renormalization group flow of the ${\rm O}(N)$ nonlinear sigma model in arbitrary dimensions.
The effective action of the model is truncated to fourth order in the derivative expansion and the flow 
is obtained by combining the non-perturbative renormalization group and the background 
field method. We investigate the flow in three dimensions and analyze the phase structure for 
arbitrary $N$. 
While a nontrivial fixed point is present in a reduced truncation of the effective action and has critical properties 
which can be related to the well-known features of the ${\rm O}(N)$ universality class, one of the fourth order operators destabilizes
this fixed point and has to be discussed carefully. 
%
%
The results about the renormalization flow of the models will serve as a reference for upcoming simulations 
with the Monte-Carlo renormalization group. 
\end{abstract}

\pacs{64.60.ae, 05.10.Cc, 11.10.Hi, 11.10.Lm} 

\maketitle

\section{Introduction}\label{introduction}
Nonlinear sigma models (NLSM) are of great interest since they appear 
as effective models for many quantum systems in various branches of physics,
ranging from solid state physics (e.g. quantum Hall effect) to particle physics (e.g. theory 
of light mesons) \cite{quHall,BrezinZinn,Delamotte,quAntiferro,Pelissetto,Zakrzewski,Novikov,Leutwyler}.
They are particularly useful in the
description of theories whose symmetries are broken below a certain scale.
From a perturbative point of view, NLSM are considered as fundamental theories only in two dimensions,
while in higher dimensions they are mainly used as effective theories, since the coupling constant 
has negative mass dimension and the theories are not perturbatively renormalizable.
However, there is the possibility that in $d=3$ dimensions
these models are non-perturbatively renormalizable, i.e. ``asymptotically safe'' \cite{weinberg}.
The concept of asymptotic safety has mainly been applied in order to provide a non-perturbative
renormalization of gravity \cite{Reuter:1996cp}. It is known that NLSM and gravity
exhibit many structural similarities, since both are described by non-polynomial interactions and have 
the same power counting behavior.
Further, both have been shown to be non-perturbatively renormalizable in a large-$N$ expansion
(see \cite{largeN} and references therein).
NLSM can therefore serve as
interesting laboratory to test and develop techniques for the more involved theory of gravity.

A promising approach to explore the non-perturbative quantization of asymptotically safe theories is 
provided by the ``functional renormalization group'' (FRG) \cite{wetterich,Reuter:1996cp}. 
It describes the scale dependence of a quantum field theory and in particular
of its effective action by means of an exact functional equation.
In order to treat theories with nontrivial target spaces like the NLSM, some advanced background techniques 
have been developed within the FRG framework, which still require a deeper understanding.
Nonlinear O($N$) models provide a useful testing ground for these conceptual issues, because
they attracted a lot of attention within statistical field theory and,
as a result, their critical properties are well-known.

The FRG approach is a 
suitable tool to investigate the phase structure of physical systems. A second order phase
transition, for instance, is related to a fixed point of the renormalization group flow. 
While we will explicitly study the nonlinear O($N$) model, it is expected that they possess
the same critical properties as their linear counterparts. The hypothesis of universality
states that two short-range theories with the same dimension and the same symmetries  
belong to the same universality class. This expectation is strongly supported by many 
explicit computations for $d=3$, see \cite{Zinn,Ballesteros,Butera,Pelissetto,Chan} and references therein.
While Monte Carlo simulations
yield the critical exponents within an O($N$)-invariant formulation, there is further 
need of non-perturbative analytical studies that are manifestly covariant and 
can confirm the critical properties of the nonlinear models.

Understanding the phase structure of field theories with curved target spaces
and advancing the techniques developed to deal with such theories within the FRG are the
main motivations for the present work. 
We will study the nonlinear O($N$)-models in a background field expansion 
\cite{Codello:2008qq,Percacci:2009fh} and compute the effective action in a truncation that 
includes all terms of fourth order in the derivatives (see also \cite{Fabbrichesi:2010xy} for some related computation).
We will derive the flow in arbitrary dimensions
and then focus on the interesting case $d=3$.
We will employ a manifestly covariant background 
field method and improve the formalism used in \cite{Percacci:2009fh}
by taking care of the specific scaling of the 
fluctuation fields in a fully non-perturbative way.

The three dimensional linear sigma model has been examined in the FRG framework for a truncation 
that contains all operators up to the fourth order in the derivative expansion \cite{Canet:2003qd}.
%
The results in \cite{Canet:2003qd} confirm the existence of a second order phase transition corresponding to a fixed point of the RG flow,
and show that the fourth order truncation provides a significant improvement over
the second order (local potential) approximation, in terms of precision with which the critical exponents
and the anomalous dimension are computed, as compared to other methods.
In our computations of the nonlinear model, the inclusion of fourth order operators also improves the sensitivity to the critical
properties in comparison to the second order truncation studied in \cite{Codello:2008qq}, such that the critical exponent of the
correlation length at the nontrivial fixed point qualitatively agrees with the expectation for the O$(N)$ universality class. 
However, one particular operator of fourth order destabilizes the system and no fixed point can be found for the full fourth order truncation.
Whether this represents a physical property of the nonlinear model is a delicate
question, since it would imply a departure of the critical behavior of the nonlinear model
from that of the linear one. It is not yet possible to give a conclusive answer to this issue, 
but we will argue that the nontrivial fixed point may reappear, once higher order 
operators are taken into account.

Another important motivation of this article is to provide explicit results for a 
direct comparison between the FRG approach and the widely used ab-initio
lattice approach to field theories. An accompanying analysis of the renormalization group 
flow of various nonlinear O($N$) models on the lattice by means of the Monte Carlo 
Renormalization Group (MCRG) method will be presented elsewhere \cite{upcoming}. 
It will be interesting to see how these two non-perturbative methods complement
each other with respect to the FP structure and critical properties.

This article is structured in the following way: Section \ref{background field method} is devoted to a covariant 
description of the O($N$) models with curved target spaces and the development of 
the background field expansion. In section \ref{functional rg} we will recall the FRG approach and 
show how to implement the background field method. Our truncation of the effective action 
is discussed in section \ref{higher derivative model and beta functions} and the running of the remaining couplings is derived. 
Section \ref{on phase transition} presents an analysis of the renormalization flow and the critical properties 
of the model. The conclusions are contained in section \ref{conclusion}.

\section{Geometry of the model}\label{background field method}

A field of the ${\rm O}(N)$ nonlinear sigma model is a map $\varphi: \Sigma \to S^{N-1}$
from spacetime $\Sigma$ to the unit-sphere in $\mathbb{R}^N$. In the present work we
will choose for $\Sigma$ the Euclidean space $\mathbb{R}^d$.
It is convenient to regard the components $\varphi^a$ as coordinates 
on the target space $S^{N-1}$. The field space of all these maps is denoted
by ${\cal M}\equiv\{\varphi:\mathbb{R}^d\to S^{N-1}\}$.

The target space is the homogeneous space $S^{N-1}={\rm O}(N)/{\rm O}(N-1)$ with ${\rm O}(N)$ 
as its isometry group. It is equipped with a unique Riemannian metric $h_{ab}$
that is ${\rm O}(N)$-invariant. The microscopic action for the model is
\begin{equation}
\label{Caction}
 S[\varphi] = \frac{1}{2}\zeta \int d^d x ~ h_{ab}(\varphi) \pmu \varphi^a \pMu \varphi^b\,,
\end{equation}
where $\zeta$ is a coupling constant.
Since the fields $\varphi^a$ have the status of coordinates,
it is natural to regard them as dimensionless, such that $\zeta$ has mass-dimension $[\zeta]=d-2$.

The action is invariant under arbitrary reparametrizations 
$\varphi\to\varphi'(\varphi)$ of the fields, provided that the metric $h_{ab}$ transforms 
as a symmetric $2$-tensor. In addition the model admits the  ${\rm O}(N)$ isometries
on the sphere as symmetries.
These isometries are generated by vector fields $K^a_i(\varphi)$ 
which satisfy a generalized angular momentum algebra,
\begin{equation}
 \left[K_i,K_j\right]=-f_{ij\ell} K_\ell\,,
\end{equation}
where $f_{ij\ell}$ are the structure constants of the Lie algebra of the rotation
group. The infinitesimal symmetries generated by the $K_i$ are nonlinear
\begin{equation}
 \label{symmetry}
 \varphi^a \to \varphi^a+\epsilon^i K^a_i(\varphi).
\end{equation}

From the invariant metric $h_{ab}$ on the sphere one obtains the
unique Levi-Civita connection $\Gamma_a{}^b{}_c$
and the corresponding Riemann tensor $R_{abcd}=h_{ac}h_{bd}-h_{ad}h_{bc}$, 
Ricci tensor $R_{ab}=(N-2)h_{ab}$ and  scalar curvature $R=(N-1)(N-2)$.

The Levi-Civita connection on the sphere is used
to construct ${\rm O}(N)$-covariant derivatives of the pull-backs
of tensors on the sphere. For example, given a pull-back of a vector on the sphere, 
its covariant derivative is
\begin{equation}\label{pull-back connection}
 \nabla_\mu v^a \equiv \partial_\mu v^a + \partial_\mu \varphi^b \Gamma_b{}^a{}_c v^c\,.
\end{equation}
The pull-back covariant derivative $\nabla$ will be used extensively throughout this work. Note that 
$\partial_\mu \varphi^a$ transform as vectors, while the coordinates $\varphi^a $ are scalars.
We also define the square of the covariant derivative 
$\Box \equiv \delta^{\mu\nu}\nabla_\mu\nabla_\nu$
and its Laplacian $ \Delta=-\Box$.

In order to construct expansions of invariant functionals like \eqref{Caction} 
such that the nonlinear symmetries \eqref{symmetry} are maintained, we will
employ a covariant background field expansion. We therefore promote the metric $h_{ab}(\varphi)$ 
on the sphere to a metric $h_{ab}(\varphi)$ on field space ${\cal M}$ where trivial 
spacetime indices have been suppressed for brevity.
In a similar manner, the Levi-Civita connection, the curvature tensors and the 
Laplacian can be promoted to ${\cal M}$ as well \cite{vilkovisky}. 
It is emphasized that the expansion variable ought to possess well-defined transformation 
properties both in the background field $\varphi^a$ and in the field $\phi^a$. 
It would for instance be a particularly hard task to construct ${\rm O}(N)$ covariant 
functionals in terms of the difference $\phi^a-\varphi^a$ of
two points in field space, as it transforms neither 
like a scalar, nor like a vector under isometries.

For $\phi$ being in a sufficiently small neighborhood of $\varphi$, there 
exists a unique geodesic in ${\cal M}$ connecting $\varphi$ and $\phi$.
We then construct the ``exponential map''
\begin{equation}\label{exponential map}
 \phi^a = {\rm Exp}_{\varphi} \xi^a = \phi^a(\varphi,\xi)\,.
\end{equation}
Here $\xi$ is an implicitly defined vector that belongs to the tangent space 
of ${\cal M}$ at $\varphi$. Also, $\xi$ is a bi-tensor, in the sense that it has 
definite transformation properties under both $\varphi^a$ and $\phi^a$ transformations
and for this reason we shall write it often as $\xi^a=\xi^a(\varphi,\phi)$.
It transforms as a vector under the ${\rm O}(N)$ transformations of $\varphi^a$ and as a scalar 
under those of $\phi^a$. By construction, the norm of $\xi$ equals the 
distance between $\varphi$ and $\phi$ in ${\cal M}$ \cite{ketov}.
The definite transformation properties make $\xi$ a candidate to parametrize any 
expansion of functionals like \eqref{Caction} around a background $\varphi$.
The correspondence is invertible, therefore in \eqref{exponential map}
we introduced the notation $\phi^a=\phi^a(\varphi,\xi)$ to be used when $\phi^a$ is understood as a function of $\xi^a$.
One always has to keep in mind that the relation between $\xi^a$ and $\phi^a$ is nonlinear.

This background field expansion can easily be applied to any functional $F[\phi]$,
\begin{equation}\label{expansion}
 F[\phi]= F[\phi(\varphi,\xi)] = F[\varphi,\xi]\,,
\end{equation}
where by abuse of notation we indicated with the same symbol the functional when 
it is a function of the single field $\phi$, 
or of the couple $\{\varphi,\xi\}$. Even though it may not be evident 
from the way the r.h.s. of \eqref{expansion} is written, a functional 
like $F[\phi]$ will never be a genuine function of two fields, but rather a function 
of the single combination $\phi^a(\varphi,\xi)$, and thus is called ``single-field'' 
functional. There may be, however, functionals of $\varphi$ and $\xi$ independently, 
and those are called ``bi-field'' functionals. As an example, we define the ``cutoff-action''
\begin{equation}\label{cutoff}
 \Delta S_k[\varphi,\xi] = \frac{1}{2} \int d^dx ~ \xi^a {\cal R}^k_{ab}(\varphi) \xi^b\,,
\end{equation}
where ${\cal R}^k_{ab}(\varphi)$ is some symmetric $2$-tensor operator constructed with $\varphi^a$.
The cutoff action will play an important role in the next section, where we will also
specify its properties. The functional \eqref{cutoff} is still invariant
under transformations of the background field $\varphi^a$ as well as the field $\phi^a$; 
however, for general ${\cal R}^k_{ab}(\varphi)$, there is no evident way to recast it as 
a functional of the single field $\phi^a$ \cite{reuter_background_works}.

We now construct an expansion of functionals like \eqref{expansion},
viewed as functions of the pair $\{\varphi,\xi\}$, in powers of $\xi$.
This can be achieved in a fully covariant way by introducing the affine 
parameter $\lambda\in [0,1]$ that parametrizes the unique geodesic connecting $\varphi$ 
and $\phi$ \cite{ketov,mukhi}. Let $\varphi_\lambda$ be this geodesic with $\varphi_0=\varphi$ and $\varphi_1=\phi$. 
Let us also define the tangent vector to the geodesic $\xi_\lambda = d\varphi_\lambda/d\lambda$ 
at the generic point $\varphi_\lambda$.
We have that \eqref{exponential map} is equivalent to
\begin{eqnarray}
 \frac{d \xi^a_\lambda}{d\lambda}
 &=&
 -\Gamma_b{}^a{}_c(\varphi_\lambda) ~ \xi^b_\lambda \xi^c_\lambda\,,
 \\
 \xi^a_0
 &=&
 \xi^a\,.
\end{eqnarray}
We can rewrite the differential equation satisfied by the tangent vectors by introducing 
the derivative along the geodesic $\nabla_\lambda\equiv \xi^a_\lambda\nabla_a$
to find that $ \nabla_\lambda \xi^a_\lambda = 0 $. By construction $\nabla_\lambda h_{ab}=0$.
Another important property is the relation to the pull-back connection
\begin{equation}
 \nabla_\lambda \partial_\mu \varphi^a_\lambda = \nabla_\mu \xi^a_\lambda\,.
\end{equation}
The commutator of the pull-back of the covariant derivative $\nabla_\lambda$ with $\nabla_\mu$ can be computed on the pull-back of a general tangent vector $v^a$
\begin{equation}
 [\nabla_{\lambda},\nabla_{\mu}]v^a = R_{cd}{}^a{}_b(\varphi_\lambda) \xi^c\partial_\mu\varphi_\lambda^d v^b\,,
\end{equation}
and is extensively needed in the covariant expansion.
Let us now use the covariant derivative along the geodesic to perform the expansion of
\eqref{expansion}. Viewing a general functional $F[\phi]$ as 
limit $\lambda\to 1$ of $F[\varphi_\lambda]$ we expand the latter in 
powers of $\lambda$ around $\lambda=0$. One can show that
\begin{equation}\label{expansion2}
 F[\phi]
 =
 \sum_{n \ge 0} \frac{1}{n!} \frac{d^n}{d\lambda^n}  \left.F[\varphi_\lambda]\right|_{\lambda=0}
 =
 \sum_{n \ge 0} \frac{1}{n!} \nabla_\lambda^n  \left.F[\varphi_\lambda]\right|_{\lambda=0}\,,
\end{equation}
where we used the fact that $F[\phi]$ is a scalar function of $\phi$.
The r.h.s. yields an expansion in powers of $\xi^a$ of the form
\begin{equation}
 F[\phi] = \sum_{n \ge 0} F^n_{(a_1,\dots,a_n)}[\varphi]\xi^{a_1}\dots\xi^{a_n}\,.
\end{equation}
As an example we give the first few terms of the expansion of the microscopic action
\eqref{Caction},
\begin{eqnarray}\label{Caction_expansion}
 S[\phi] &=&
 S[\varphi]
 + \zeta \int \! d^dx ~ h_{ab}\partial_\mu \varphi^a\nabla^\mu \xi^b
 \nonumber
 \\
 &
 +& \frac{\zeta}{2} \int \! d^dx ~ \Bigl(\nabla_\mu \xi^a\nabla^\mu \xi_a + R_{abcd}\partial_\mu\varphi^b\partial^\mu\varphi^c\xi^a\xi^d\Bigr)
 \nonumber
 \\
 &
 +& {\cal O}(\xi^3)\,.
\end{eqnarray}

\section{Functional RG}\label{functional rg}

We define the scale-dependent average effective action of the nonlinear ${\rm O}(N)$ 
sigma model \cite{Codello:2008qq,Percacci:2009fh} by the functional integral
\begin{eqnarray}\label{erge1}
 e^{-\Gamma_k[\varphi,\bar{\xi}]} &=& \int \! D\xi \, \mu[\varphi] \, e^{-S[\phi]+\frac{\delta\Gamma_k}{\delta \bar{\xi}^a}[\varphi,\bar{\xi}] (\bar{\xi}^a-\xi^a)}\nonumber\\
 &&\hskip1.7cm\times\, e^{-\Delta S_k[\varphi,\bar{\xi}-\xi]}\,,
\end{eqnarray}
with density of the covariant measure
$\mu[\varphi] = {\rm Det}\, h(\varphi)^{1/2}
$.
Averages are defined through the path integral on the r.h.s. of \eqref{erge1}, 
i.e. $\bar{\xi}^a\equiv\left<\xi^a\right> $. Using the relation 
$\bar{\xi}^a = \xi^a(\varphi,\bar{\phi})$ one can obtain $\bar{\phi}^a$ that plays the 
role of full average field of the model. The definition \eqref{erge1} differs from the 
usual definitions of effective action only by the presence of the functional $\Delta S_k$ 
defined  in \eqref{cutoff}. This cutoff action is chosen in such a way that the 
average effective action $\Gamma_k[\varphi,\bar{\xi}]$ interpolates between
the classical action at the scale $k=\infty$ and the full effective action at $k=0$.
It is chosen to be quadratic in $\xi^a$ with a $\varphi$-dependent kernel ${\cal R}^k_{ab}$
that regularizes the infrared contributions of the fluctuations $\xi$. This kernel will be
called ``cutoff kernel'' from now on. The reference scale $k$ distinguishes between 
infrared energy scales $k'\lesssim k$ and ultraviolet ones $k'\gtrsim k$ and the cutoff action 
leads to a scale dependent effective action in the following way:
The propagation of infrared modes in the path integral \eqref{erge1} is suppressed,
such that only the ultraviolet modes are integrated out and we are left with
a scale-dependent average effective action $\Gamma_k$ for the remaining
infrared modes. In order to provide for an interpolation between 
the classical and the effective action, the cutoff-kernel has to fulfill the two conditions
$\lim_{k\rightarrow 0}{\cal R}^{k}_{ab}[\varphi]=0$ and $\lim_{k\rightarrow \infty}{\cal R}^{k}_{ab}[\varphi]=\infty$.

Due to the presence of the cutoff term, $\Gamma_k[\varphi,\bar{\xi}]$ 
is genuinely a bi-field functional \cite{pawlowski,reuter_background_works},
\begin{equation}
 \hat{\Gamma}_k[\varphi,\bar{\phi}] = \Gamma_k[\varphi,\bar{\xi}(\varphi,\bar{\phi})]\,.
\end{equation}
This observation is important in order to understand that the only way to construct a single field  
effective action is to set $\varphi=\bar{\phi}$ or equivalently $\bar{\xi}=0$ 
and consider
\begin{equation}
 \bar{\Gamma}_k[\bar{\phi}] = \hat{\Gamma}_k[\bar{\phi},\bar{\phi}] = \Gamma_k[\bar{\phi},0]\,.
\end{equation}
The limit $k\to 0$ of $\bar{\Gamma}_k[\bar{\phi}]$ coincides with the well known effective action 
introduced by deWitt \cite{dewitt}.

The very useful feature of the definition \eqref{erge1} is that $\Gamma_k[\varphi,\bar{\xi}]$ 
satisfies a functional renormalization group equation \cite{wetterich}
\begin{equation}\label{erge2}
k\partial_k \Gamma_k[\varphi,\bar{\xi}] = \frac{1}{2}{\rm Tr}\left(
\frac{k\partial_k{\cal R}_k[\varphi]}
{\Gamma_k^{(0,2)}[\varphi,\bar{\xi}]+{\cal R}_k[\varphi])}\right)\,.
\end{equation}
The functional differential equation \eqref{erge2} is equivalent to the definition \eqref{erge1}
once an initial condition $\Gamma_\Lambda[\varphi,\bar{\xi}]$ for some big UV-scale 
$\Lambda$ is specified, which accounts for the renormalization
prescription and the inclusion of counter terms.

Having an equation like \eqref{erge2} at our disposal, it is possible to investigate 
properties of the quantum field theory under consideration, without having to explicitly 
compute the path integral \eqref{erge1}. In particular we can investigate whether 
or not the theory admits a second order phase transition for some value of its coupling 
constants. It is well known that, from a renormalization group perspective, the critical
behavior is dictated by a the fixed points of the renormalization group flow.

\section{Higher derivative model and beta functions}\label{higher derivative model and beta functions}

It has been stressed in the previous section that the flow equation \eqref{erge2} is 
non-perturbative in nature such that we can explore nontrivial 
features of the renormalization group flow. For this purpose  
we must find an ansatz for the scale-dependent effective action 
that includes all relevant operators.
We use a covariant expansion of the effective action in orders of derivatives and
take into account all possible operators containing up to four derivatives. 
We furthermore consider the split
\begin{eqnarray}\label{split}
 \Gamma_k[\varphi,\xi] &=& \Gamma^{\rm s}_k[\phi(\varphi,\xi)]+\Gamma^{\rm b}_k[\varphi,\xi]\,,
\end{eqnarray}
where we dropped the overline on the arguments $\xi$ and $\phi$ for notational simplicity.
Some comments are in order: We have already stressed that due to the presence of the 
cutoff action in the definition \eqref{erge1} the functional $\Gamma_k[\varphi,\xi]$ depends
on the two fields $\{\varphi,\xi\}$ separately, and not only on the combination $\phi(\varphi,\xi)$.
In the split \eqref{split} we collected the contribution that can actually be written as functional 
of the ``full'' field $\phi$ into $\Gamma^{\rm s}_k[\phi(\varphi,\xi)]$, and parametrized 
the rest as $\Gamma^{\rm b}_k[\varphi,\xi]$, 
with $\Gamma^{\rm b}_k[\varphi,0]=0$ \cite{reuter_background_works}.

As an ansatz for $\Gamma^{\rm s}_k[\phi(\varphi,\xi)]$, we use the most general local action 
up to fourth order in the derivatives \cite{Percacci:2009fh,bk}
\begin{eqnarray}\label{Eaction1}
 \Gamma^{\rm s}_k[\phi]
 &=&
 \frac{1}{2} \int d^d x~\Bigl(
 \zeta_k h_{ab} \pmu \phi^a \pMu \phi^b
 + \alpha_k h_{ab} \Box \phi^a \Box \phi^b
 \nonumber\\
 &&
 + T_{abcd}(\phi) \pmu \phi^a \pMu \phi^b  \pnu \phi^c \pNu \phi^d\Bigr)\,.
\end{eqnarray}
The isometries of the model are respected only if the tensor $T_{abcd}$ is invariant.
In the simple case of the ${\rm O}(N)$-model, there exists a unique (up to normalization)
invariant $2$-tensor $h_{ab}$ and all higher rank invariant tensors are constructed 
from $h_{ab}$. According to the symmetries $T_{abcd}=T_{((ab)(cd))}$ that can be 
deduced trivially from \eqref{Eaction1}, we see that the most general 
parametrization of $T_{abcd}$ reads
\begin{eqnarray}\label{Ttensor}
 T_{abcd}
 &=&
 L_{1,k} h_{a(c}h_{d)b} +L_{2,k} h_{ab}h_{cd}\,.
\end{eqnarray}
Using this parametrization in \eqref{Eaction1}, we realize that a total of four 
couplings has been introduced $\{\zeta_k,\alpha_k,L_{1,k},L_{2,k}\}$.
They parametrize the set of operators that we include in our truncation 
and encode the explicit $k$-dependence of $\Gamma^{\rm s}_k$.

Now we need a consistent ansatz for $\Gamma^{\rm b}_k[\varphi,\xi]$.
It is important to dress the $2$-point function of the field $\xi^a$ correctly, 
because it is the second derivative w.r.t. $\xi$ which determines the flow \eqref{erge2}.
We choose
\begin{eqnarray}\label{Eaction2}
 \Gamma^{\rm b}_k[\varphi,\xi]
 &=&
 \Gamma^{\rm s}_k[\phi(\varphi,Z_k^{1/2}\xi)]-\Gamma^{\rm s}_k[\phi(\varphi,\xi)]
 \nonumber\\
 &&
 +Z_k\frac{m_k^2}{2}\int d^dx ~ h_{ab}\xi^a\xi^b\,.
\end{eqnarray}
This choice includes a mass term for the fluctuation fields as well as a nontrivial 
wave function renormalization of these fields $\xi^a \rightarrow Z_k^{1/2} \xi^a$,
which takes into account the possibility that the fields $\varphi^a$ and $\xi^a$ 
may have a different scaling behavior. In a first step beyond the covariant
gradient expansion we assume that the wave function renormalization only depends
on the scale $k$. Later we shall see that the wave function renormalization
enters the flow solely via the anomalous dimension $\eta_k=-k\partial_kZ_k/Z_k$ of the 
fluctuation field.
Contrary to $Z_k$ the square mass $m^2_k$ enters the flow equation directly.
It is the most direct manifestation of the fact that $\Gamma_k[\varphi,\xi]$ is a function of the two variables separately.
While one could add many other covariant operators to $\Gamma^b_k$,
we first want to investigate the effects of these simple structures.
Besides the different expansion \eqref{rhsExp} of the flow equation, the ansatz for $\Gamma^{\rm b}_k[\varphi,\xi]$ represents the main conceptual departure of our computation from the one in \cite{Percacci:2009fh}, where the approximation
$Z_k=1$ and $m_k^2=0$ was employed. It will become clear in the following that
the fields $\varphi^\alpha$ and $\xi^\alpha$ possess rather different wavefunction renormalizations,
therefore making the inclusion of the relative factor $Z_k^{1/2}$ a necessary ingredient for a consistent truncation.

Now we can plug the ansatz \eqref{split} into the flow equation \eqref{erge2}.
Projecting the r.h.s. of the flow equation onto the same operators that appear in 
the ansatz for $\Gamma_k$ it is possible to determine the non-perturbative beta 
functions of the model. In order to proceed we must specify the cutoff kernel
appearing in \eqref{cutoff}. We want it to be a function of $\varphi$ solely 
through the Laplacian and it should otherwise be proportional to the metric $h_{ab}$
\begin{eqnarray}\label{cutoff profile}
 {\cal R}^k_{ab}[\varphi] &=& Z_k h_{ab} R_k[\Delta]\,.
\end{eqnarray}
The cutoff is specified through the non-negative ``profile function'' $R_k[z]$ which
can be regarded as a momentum-dependent mass. The choice \eqref{cutoff profile} 
ensures that we are coarse-graining  the theory relative to the modes of 
the covariant Laplacian $\Delta$. In order to serve as a good  cutoff the 
function $R_k[z]$ has to be monotonic 
in the variable $z$ and such that $R_k[z]\simeq 0$ for $ z \gtrsim k^2$.
The wave function renormalization of the $\xi^a$ field has been used in 
\eqref{cutoff profile} as an overall parametrization.
It is convenient to compute the scale derivative of \eqref{cutoff profile} 
already at this stage. We obtain
\begin{eqnarray}\label{cutoff profile derivative}
 k\partial_k{\cal R}^k_{ab}[\varphi] &=& Z_k h_{ab} \left(k\partial_kR_k[\Delta]-\eta R_k[\Delta]\right)\,.
\end{eqnarray}
For the sake of convenience we will compute the beta functions of the two sets of couplings $\{\zeta_k,\alpha_k,L_{1,k},L_{2,k}\}$ and $\{Z_k,m^2_k\}$
in two separate steps.
We begin the computation of the flow of $\Gamma^{\rm s}_k[\phi(\varphi,\xi)]$ by considering 
the limit $\xi\to 0$ of \eqref{erge2}. The result is a flow equation of the form
\begin{eqnarray}\label{erge background}
 k\partial_k \Gamma^{\rm s}_k[\varphi]
 &=&
 \frac{1}{2}{\rm Tr}\left(\frac{k\partial_k{\cal R}_k[\varphi]}{
 \Gamma_k^{(0,2)}[\varphi,0]+{\cal R}_k[\varphi]}\right)
 \nonumber\\
 &=&
 \frac{1}{2}{\rm Tr}\,\{G_k{}^a{}_a \left(k\partial_kR_k[\Delta]-\eta R_k[\Delta]\right)\}
 \,,
\end{eqnarray}
where the dependence on $Z_k$ enters via the modified propagator 
$ G_k{}^{ab}$ which is the inverse of $(Z_k^{-1}\Gamma_k^{(0,2)}[\varphi,0] +{\cal R}_k[\Delta])_{ab}$.
It shows how the fluctuations $\xi$ drive the flow 
of the couplings $\{\zeta_k,\alpha_k,L_{1,k},L_{2,k}\}$.
The modified propagator is computed from \eqref{split} using
(\ref{Eaction1},\ref{Eaction2}) and reads
\begin{eqnarray}\label{modified propagator}
 G_k
 &=& \left(P_k[\Delta] +\Sigma \right)^{-1}
 \\
 P_k[\Delta]
 &=&
 \alpha_k \Delta^2 + \zeta_k \Delta + m^2 \mathbb{I} + R_k[\Delta]
 \\
 \Sigma
 &=&
 B^{\mu\nu} \nabla_\mu \nabla_\nu + C^\mu\nabla_\mu + D\,,
\end{eqnarray}
where indices in the tangent space to ${\rm O}(N)$ have been suppressed for brevity.
The matrices $B^{\mu\nu}$, $C^\mu$ and $D$ are endomorphisms in the tangent
space. In the following we will need the explicit form of two of them:
\begin{eqnarray}\label{operators}
 B_{ab}^{\mu\nu}
 &=&
 2 \delta^{\mu\nu} (\alpha_k R_{acbd}- T_{abcd}) \partial_{\rho} \varphi^c \partial^{\rho} \varphi^d\nonumber\\
 &&- 4 T_{acbd} \pMu \varphi^c \pNu \varphi^d
 \nonumber\\
 D_{ab}
 &=&
 -\zeta_k R_{acbd}\partial_{\rho} \varphi^c \partial^{\rho} \varphi^d - \alpha_k R_{acbd} \Box \varphi^c \Box \varphi^d \nonumber\\
 && + (\alpha_k R_{acde}R_{bfg}{}^{e}+2R_{e(ab)f}T^e{}_{gcd})\nonumber\\
 && ~ \times \partial_{\rho} \varphi^c \partial^{\rho} \varphi^d \partial_{\sigma}\varphi^f \partial^{\sigma}\varphi^g \,.
\end{eqnarray}
Each term in $B^{\mu\nu}$ and $D$ consists of at least two derivatives of the field $\varphi^a$.
This implies that a Taylor expansion of \eqref{modified propagator} in $\Sigma$ is possible,
because in our truncation ansatz we are interested only in terms up to fourth order in derivatives \cite{rgmachine}.
The tensor $C^\mu$ contains three derivatives of $\varphi^a$ and thus
can be ignored in our truncation. Thus the expansion reads
\begin{equation}
\label{rhsExp}
 G_k \!= \! P_k^{-1}\! - \!P_k^{-1}\Sigma P_k^{-1}\! +\! P_k^{-1}\Sigma P_k^{-1}\Sigma P_k^{-1}\! +{\cal O}(\partial^6)\,.
\end{equation}
Inserting this expansion into \eqref{erge background} and using the cyclicity of the trace we obtain
\begin{eqnarray}\label{erge background expansion}
 k\partial_k \Gamma^{\rm s}_k[\varphi]
 &=&
 \frac{1}{2}{\rm Tr}\,f_1[\Delta]
 -\frac{1}{2}{\rm Tr}\,\Sigma f_2[\Delta]
 \nonumber\\
 &&
 +\frac{1}{2}{\rm Tr}\,\Sigma^2 f_3[\Delta]
 +{\cal O}(\partial^6)
 \,,
\end{eqnarray}
where we defined the functions
\begin{eqnarray}
 f_l[z]\equiv \frac{k\partial_kR_k[z]-\eta R_k[z]}{P_k[z]^l}\,.
\end{eqnarray}
In \eqref{erge background expansion} we also used the fact that we can 
commute $\Sigma$ and $P_k[\Delta]$ in the third term in the expansion,
because their commutator leads to terms of order ${\cal O}(\partial^6)$. 

The traces appearing in \eqref{erge background expansion} are computed using 
off-diagonal heat kernel methods \cite{off-diagonal}. To outline the general 
procedure briefly we consider the traces 
\begin{equation}\label{general trace}
 {\rm Tr}\,(\nabla_{\mu_1}\dots\nabla_{\mu_r} f[\Delta])\,
\end{equation}
which transform as tensors under isometries.
We are interested in the particular cases $0\leq r\leq 4$ and $f[\Delta]=f_l[\Delta]$ 
for some $l=1,2,3$. Introducing the inverse Laplace transform ${\cal L}^{-1}[f](s) $ of $f[z]$,
we rewrite \eqref{general trace} as
\begin{equation}\label{heat kernel}
 \int_0^{\infty} ds\,{\cal L}^{-1}[f](s) \, {\rm Tr}\,(\nabla_{\mu_1}\dots\nabla_{\mu_r} e^{-s \Delta})\,.
\end{equation}
The expansion of the trace $ {\rm Tr}\,(\nabla_{\mu_1}\dots\nabla_{\mu_r} e^{-s \Delta})$
is obtained from the off-diagonal heat kernel expansion (the case $r=0$ yields the trace 
of the heat kernel itself). This expansion is an asymptotic small-$s$ expansion 
that corresponds, for dimensional and covariance reasons, to an expansion
in powers of the curvature and  covariant derivative. Thus we have
\begin{eqnarray}\label{heat kernel expansion}
 {\rm Tr}\,(\nabla_{\mu_1}\dots\nabla_{\mu_r} e^{-s \Delta})
 =
 \sum_{n=0}^\infty \frac{B_{\mu_1\dots\mu_r ,n}}{(4\pi s)^{d/2}}s^{\frac{2n-[r]}{2}}\,,
\end{eqnarray}
where $[r]=r$ if $r$ is even and $[r]=r-1$ if $r$ is odd.
The coefficients $B_{\mu_1\dots\mu_r,n} $ contain a number of powers of the derivatives of the field that increases with $n$,
thus only a finite number of them is needed to compute the traces with ${\cal O}(\partial^4)$ accuracy \cite{off-diagonal}.
The final step in evaluating \eqref{general trace} is the s-integration. We define
\begin{eqnarray}
 Q_{n,l} = \frac{1}{(4\pi)^{d/2}}\int_0^{\infty}ds\, s^{-n} {\cal L}^{-1}[f_l](s)\,,
\end{eqnarray}
which we will denote as ``$Q$-functionals'' that can be rewritten (for positive $n$) as Mellin transforms of $f_l(z)$
\begin{eqnarray}\label{Qfunctionals}
 Q_{n,l} = \frac{1}{(4\pi)^{d/2}\Gamma[n]}\int_0^{\infty} dz\, z^{n-1} f_l[z].
\end{eqnarray}
In the heat kernel expansion, the $Q$-functionals play a similar role that is played by the regularized
Feynman diagrams in perturbation theory.
With the help of the heat kernel method we now expand the right hand side in
the flow equation \eqref{erge background expansion} and obtain
\begin{align}
\label{erge background expansion2}
 k\partial_k\Gamma_k^{\rm s}[\varphi]
 =& 
 \frac{1}{2}{\rm tr}\! \int \! d^d x \Bigl\{
 \frac{1}{12} Q_{\frac{d}{2}-2,1} \left[\nabla_\mu,\nabla_\nu\right]^2
 \nonumber\\
 &+ \frac{1}{2} Q_{\frac{d}{2}+1,2} B^{\mu}{}_{\mu}
 - \frac{1}{2} Q_{\frac{d}{2},2} B^{\mu\nu}\left[\nabla_\mu,\nabla_\nu\right]
 \nonumber\\
 &+ \frac{1}{2}Q_{\frac{d}{2}+2,3}\Bigl(B^{(\mu\nu)}B_{\mu\nu}+\frac{1}{2}(B^{\mu}{}_{\mu})^2\Bigr)
 \\
 &- Q_{\frac{d}{2},2} D - Q_{\frac{d}{2}+1,3}B^{\mu}{}_{\mu} D
 + Q_{\frac{d}{2},3} D^2  \Bigr\}\,\nonumber
\end{align}
whereby the tensor fields $B$ and $D$ are given in \eqref{operators}.
The beta functions for $\{\zeta_k,\alpha_k,L_{1,k},L_{2,k}\}$ are defined
as their log-$k$-derivatives and are denoted by $\{\beta_\zeta,\beta_\alpha,\beta_{L_1},\beta_{L_2}\}$.
We extract them by evaluating both sides of the flow equation at $\xi = 0$.
With \eqref{Eaction1} the left hand side reads
\begin{eqnarray}\label{lhs flow background}
 k\partial_k \Gamma^{\rm s}_k[\varphi]
 &=&
 \frac{1}{2} \int d^d x~\Bigl(
 \beta_\zeta \pmu \varphi^a \pMu \varphi_a
 + \beta_\alpha \Box \varphi^a \Box \varphi_a
 \nonumber\\
 &
 & \hskip-10mm +\,\beta_{L_1} (\pmu \varphi^a \pMu \varphi^b)^2 + \beta_{L_2} (\pmu \varphi^a \pMu \varphi_a)^2
 \Bigr)\,,
\end{eqnarray}
and a comparison with the right hand side in
\eqref{erge background expansion2} yields the beta functions
\begin{align}\label{beta functions}
 &\beta_\zeta
 =\zeta_k (N-2) Q_{\frac{d}{2},2} +(N-2) d \alpha_k Q_{\frac{d}{2}+1,2}
 \nonumber\\
 &~~ + L_{1,k}(N+d)Q_{\frac{d}{2}+1,2}
 - L_{2,k}((N-1)d+2)Q_{\frac{d}{2}+1,2}
 \nonumber\\
 &\beta_\alpha
 =(N-2) Q_{\frac{d}{2},2} \alpha_k
 \,.
\end{align}
The remaining beta functions $\beta_{L_1}$ and $\beta_{L_2}$ are quite long and given in the appendix \eqref{bL}.
It suffices to know that
$\beta_{L_1},\,\beta_{L_2}$ are quadratic polynomials of $L_1$ and $L_2$
and that the coefficients of these polynomials depend
nontrivially on both the parameters $N$ and $d$, and on the other two couplings.
The combined limit $\{d=4,\zeta_k= 0,\eta=0,m^2_k=0\}$ of $\{\beta_{L_1},\beta_{L_2}\}$
is known to yield a universal result which can be compared with the results of 
chiral perturbation theory in the case $ S^3\simeq {\rm SU(2)}$ \cite{Percacci:2009fh,hasenfratz}.
We checked that we can reproduce the perturbative results in this limit.
The results of \cite{Percacci:2009fh} are recovered fully in the limit $d=4$ and for general $S^{N-1}$,
if a chiral perturbative expansion is performed to $1$-loop.

Let us finally outline the computation of the flow of $Z_k$ and $m^2_k$.
The simplest setting to perform this is a vertex expansion of the flow \eqref{erge2} 
in powers of the field $\xi^a$. We first notice that for a constant background 
field $\varphi^a_c$ the ansatz for the effective action \eqref{split} reduces to
\begin{align}\label{fluctuations action}
 \Gamma_k[\varphi_c,\xi]
 =&
 \frac{Z_k}{2}\int d^dx \Bigl\{\zeta_k h_{ab} \nabla_\mu\xi^a\nabla^\mu\xi^b + \alpha_k h_{ab} \Box\xi^a\Box\xi^b
 \nonumber\\
 &
 + m_k^2 h_{ab} \xi^a \xi^b
 +\frac{1}{3} \zeta_k Z_k R_{abcd}\xi^a \xi^d \Pmu\xi^c \PMu \xi^b
 \nonumber\\
 &
 +\frac{4}{3} \alpha_k Z_k R_{abcd} \xi^a \Pmu \xi^b \PMu \xi^d \Box \xi^c
 \nonumber\\
 &
 + \frac{1}{3} \alpha_k Z_k R_{abcd} \xi^a \Box \xi^b \Box \xi^c \xi^d \\
 &
 +  Z_k T_{abcd} \Pmu \xi^a \PMu \xi^b \Pnu \xi^c \PNu \xi^d
 \Bigr\} + {\cal O}(\xi^6)\,.\nonumber
\end{align}
In this particular limit the pull-back connection \eqref{pull-back connection}
becomes trivial: $\nabla_\mu=\partial_\mu$ and $\Box=\partial^2$.
This observation is particularly useful, since we can now easily perform the 
computations in momentum space. We define $ \xi^a(x) = \int d^dq\, e^{\imath q x} \xi^a_q$ 
and obtain from \eqref{fluctuations action} the $2$-point function for incoming momentum $p^\mu$:
\begin{align}
 \Gamma^{(0,2)}_k[\varphi_c,0]_{p,-p}
 &=
 Z_k (\alpha_k p^4 + \zeta_k p^2 +m_k^2)\,.
\end{align}
We also compute its scale derivative
\begin{eqnarray}\label{two point function flow}
 k\partial_k\Gamma^{(0,2)}_k[\varphi_c,0]_{p,-p}
 &=&
 Z_k \Bigl((\beta_\alpha -\eta \alpha_k) p^4\\ 
 &&\hskip-.5cm +\, (\beta_\zeta-\eta \zeta_k) p^2
 +(\beta_{m^2} -\eta m_k^2) \Bigr)\,.\nonumber
\end{eqnarray}
On the other hand, the quantity $k\partial_k\Gamma^{(0,2)}_k[\varphi_c,0]_{p,-p}$
can be computed from \eqref{erge2} by applying two functional derivatives w.r.t. $\xi^a$,
taking the limit $\varphi^a=\varphi^a_c={\rm const.}$ and transforming to momentum space.
After these manipulations, the flow equation \eqref{erge2} reduces to
\begin{eqnarray} \label{two point function flow rhs}
 k\partial_k\Gamma_k^{(0,2)}[\varphi_c,0]_{p,-p}&&\\
 &&\hskip-2cm =
 -\frac{1}{2 Z_k} {\rm Tr} f_2(q^2) \Gamma_k^{(0,4)}[\varphi_c,0]_{p,-p,q,-q}\,.\nonumber
\end{eqnarray}
The momentum space $4$-point vertex function $\Gamma_k^{(0,4)}[\varphi_c,0]$ is obtained 
from \eqref{fluctuations action} 
and has to be traced over two of its four indices, while the $3$-point function vanishes 
for $\varphi^a={\rm const.}$ therefore playing no role in our computation.
The trace that appears in \eqref{two point function flow rhs} 
consists of an internal trace on the tangent space of the model, that involves two of the 
four indices of the $4$-vertex, and a momentum space integral $\int d^dq/(2\pi)^d$. The 
final result is a long expression that depends solely on $p^2$. Comparing the power $p^n$ 
with $n=0,2,4$ of \eqref{two point function flow rhs} with those of \eqref{two point function flow} 
and dividing both sides by $Z_k$, we can determine the coefficients
\begin{align}\label{fluctuations system}
 \beta_\alpha -\eta \alpha_k
 &= 
 \frac {1}{3}(N-2)\alpha_k  Q_{\frac{d}{2},2}
\nonumber \\
 \beta_\zeta\,-\eta \zeta_k
\,&= 
 \frac{1}{3} (N-2) \zeta_k  Q_{\frac{d}{2},2}
  -\Bigl((dN-d+2) L_{2,k} 
  \nonumber \\
 &~~~  +(N+d) L_{1,k}  
  - (N-2) d \alpha_k\Bigr) Q_{\frac{d}{2}+1,2}
\nonumber \\
 \beta_{m^2} -\eta m_k^2
 \,&=  
\frac{1}{12}(N-2)d(d+2) \alpha_k Q_{\frac{d}{2}+2,2} 
\nonumber \\
 & ~~~ +\frac{1}{6}(N-2)d \zeta_k Q_{\frac{d}{2}+1,2}\,.
\end{align}
As anticipated, there is no explicit dependence on $Z_k$, because it is a redundant parameter.

One interesting feature of the method arises at this point:
Using \eqref{beta functions}, we can solve the system of equations \eqref{fluctuations system} 
in terms of the two unknown quantities $\{\eta,\beta_{m^2}\}$.
For a solution to exist, one equation of \eqref{fluctuations system} must be 
redundant and it is a nontrivial check of our computation, at this stage, that this 
actually holds true. The final result for the anomalous scaling reads 
\begin{align}\label{anomalous scaling}
 &\eta=\frac{2}{3}(N-2) Q_{\frac{d}{2},2}\,.
\end{align}

\section{${\rm O}(N)$ phase diagram}\label{on phase transition}

We will now analyze in more detail the structure of the $\beta$-functions and 
the resulting phase diagram. For this purpose we focus on three spacetime dimensions as a 
particularly interesting case which has been intensively studied \cite{Pelissetto,Ballesteros,Butera}. 
Concerning the two dimensional case, we just want to mention that our computation 
reproduces the well-known statement that the theory has no nontrivial fixed point, but rather is 
asymptotically free \cite{Polyakov}.

In order to evaluate the Q-functionals and hence the explicit running of the couplings 
in three dimensions, we have to choose a specific regulator \eqref{cutoff profile} that fulfills the 
requirements described in section \ref{functional rg}. We decided to choose a 
modified ``optimized cutoff'' \cite{OptReg}:
\begin{align} \label{optimized}
  R_k[z] = \left(\zeta_k (k^2-z)+ \alpha_k (k^4-z^2)\right)\Theta(k^2-z) 
\end{align}
Note that the $k$-subscript of the couplings will be suppressed in the following. This choice of regulator enables us to calculate the Q-functionals explicitly,
\begin{eqnarray}\label{Q-optimized}
Q_{n,l} &=& \frac{k^{2n+2}}{(4\pi)^{d/2}\,\Gamma(n)}\,\Bigl(
\frac{ (2n+2-\eta+\partial_t)\zeta }{n(n+1)(\zeta k^2 + \alpha k^4 + m^2)^l}\nonumber\\
&&\hskip10mm+\frac{2 k^2 (2 n + 4 -
\eta+\partial_t)\alpha}{n(n+2)(\zeta k^2 + \alpha k^4 + m^2)^l} \Bigr) \,,
\end{eqnarray}
but it renders the system of differential equations rather involved, since 
the derivatives $\partial_t \alpha\equiv k\partial_k\alpha = \beta_{\alpha}$ 
and $\partial_t \zeta = \beta_{\zeta}$ also appear on the r.h.s. of the flow equation.
We note that the $Q$-functionals \eqref{Q-optimized} possess a threshold-like structure
due to the presence of the mass $m^2$. Further, they are linear in the anomalous scaling $\eta$ and in the beta functions.

Under the condition \eqref{Q-optimized}, the systems \eqref{beta functions} and \eqref{fluctuations system}
contain the beta functions of the theory in an implicit form.
In order to determine their explicit form it is necessary to solve together the two systems
in terms of the quantities $\{\beta_\zeta,\beta_\alpha,\beta_{L_1},\beta_{L_2},\eta,\beta_{m^2}\}$.
Fortunately, the joint system is linear in these quantities, because the $Q$-functionals are.
As a final step we rewrite the result in terms of the dimensionless couplings 
$\tilde{\zeta} = k^{2-d} \zeta$, $\tilde{\alpha}=k^{4-d}\alpha$, $\tilde{L}_1=k^{4-d}L_1$, $\tilde{L}_2=k^{4-d}L_2$ and $\tilde{m}^2=k^{-d}m^2$.
and obtain the beta functions $\{\beta_{\tilde{\zeta}},\beta_{\tilde{\alpha}},\beta_{\tilde{L}_1},
\beta_{\tilde{L}_2},\beta_{\tilde{m}^2}\}$, which are involved rational functions and hence not given here explicitly.

Now we are ready to study the phase diagrams and the critical properties that arise from 
these flow equations. We will proceed in a systematic way, by including more and more operators 
in our truncation. The simplest truncation that only contains the coupling $\zeta$ was already 
studied in \cite{Codello:2008qq}. Their work points to the existence of a 
nontrivial fixed point in dimensions larger than two and hence to the possibility of
non-perturbative renormalizability. 

\begin{figure}
\includegraphics[width=0.48\textwidth]{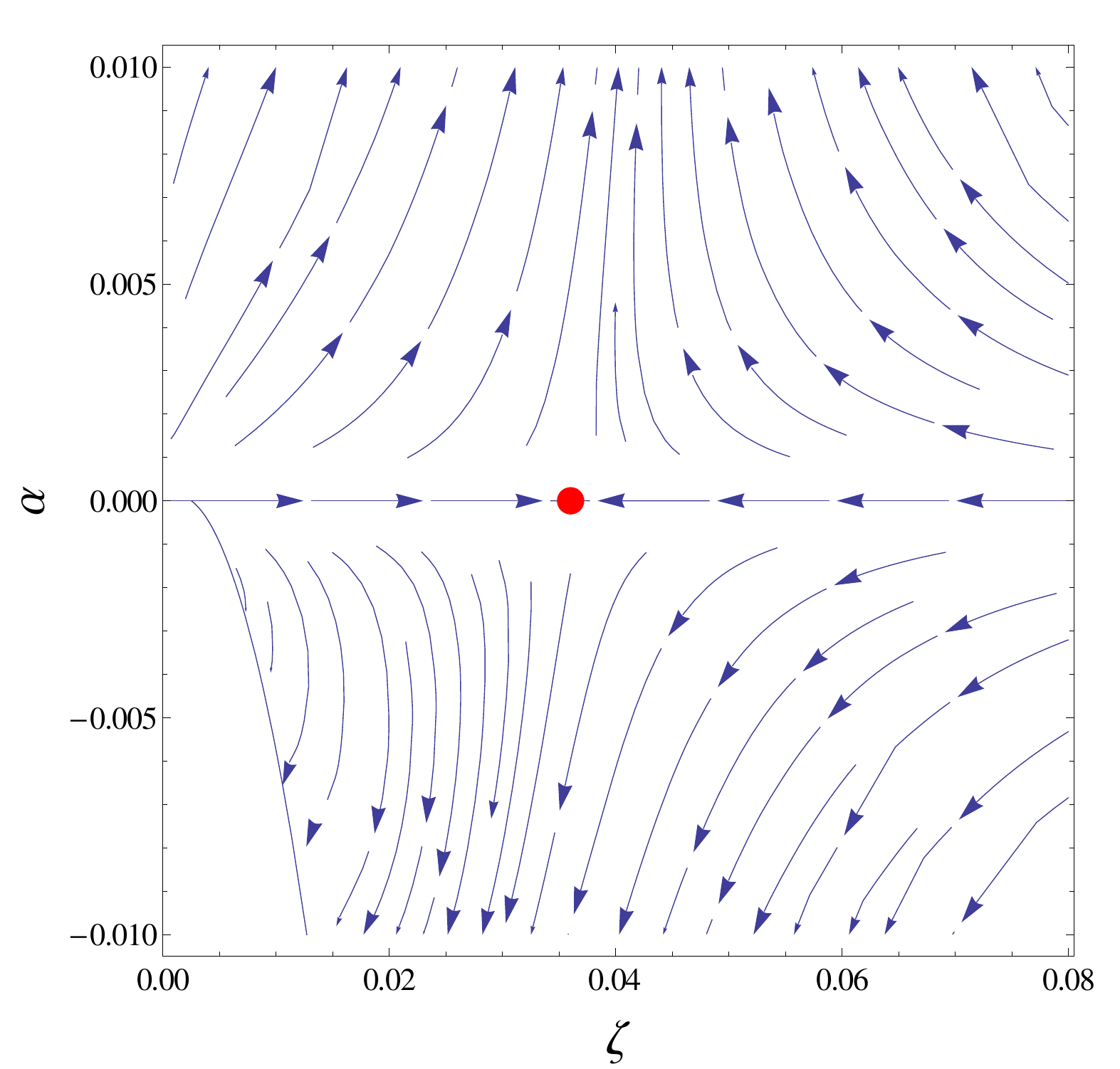}
\caption{The flow of the couplings for the truncation with two 
couplings $\alpha$ and $\zeta$ for $N=3$. The arrows point toward the 
ultraviolet. The removed region lies beyond an unphysical singularity
which is introduced by the choice of the cutoff.
}
\label{alphaflow}
\end{figure}

We want to add the coupling $\alpha$ and the related fourth-order operator. The corresponding
renormalization group flow is depicted in Figure \ref{alphaflow} (for the case $N=3$) and confirms the
nontrivial fixed point found in the simpler truncation. The critical couplings 
are $\tilde{\zeta}^* = 16(N-2)/(45\pi^2)$ and $\tilde{\alpha}^*=0$.
Note that the arrows of the flow point into the direction of increasing $k$, i.e. 
towards the ultraviolet. It is interesting to note that the coupling $\alpha$ belongs to 
an IR-irrelevant operator and vanishes at the fixed point. In fact, already 
the simple structure of $\beta_{\alpha}$ that is obtained from \eqref{beta functions} reveals that $\alpha$ 
has to vanish for every possible fixed point, as the flow of the dimensionless 
coupling $\tilde{\alpha}$ reads in $d=3$:
\begin{align}
 \beta_{\tilde{\alpha}} = \tilde{\alpha} + (N-2)~Q_{\frac{3}{2},2}~\tilde{\alpha}
\end{align}
Since $Q_{\frac{3}{2},2}$ is strictly positive for any reasonable regulator, the only 
possible fixed point value is $\tilde{\alpha}=0$. This statement remains true when
we include the couplings $L_1,L_2$.
Further, it is a general feature of the system \eqref{beta functions}
that at any FP for which $\tilde{\alpha}=0$, the critical $\alpha$ direction in phase space is decoupled from the others
and therefore cannot contribute significantly to the critical properties of the other couplings.
There is
also a fixed point for 
$\lambda = 1/\alpha = 0$, but this is a trivial one whose critical exponents are equal 
the canonical mass dimensions of the operators. It is the three-dimensional analogue
of the fixed point in four dimensions which is discussed in \cite{Percacci:2009fh}.

The result $\tilde{\alpha}^*=0$ agrees with an alternative computation of the effective 
action of the nonlinear ${\rm O}(N)$-model up to fourth order, which is presented in \cite{Chan}, and 
in which a term  $\propto \partial^2\phi\partial^2\phi$ is not generated, either.
However, it is very likely that an extension of the truncation to the sixth order 
in derivatives and an inclusion of operators  like e.g. $\Box \phi_a \Box \phi^a \partial_{\mu} \phi^b 
\partial^{\mu}\phi_b$ will affect the running of $\alpha$ and shift the position of the fixed point.
This discussion will be relevant when we compare our results with the renormalization flows 
obtained by the Monte Carlo Renormalization Group (MCRG), that will be studied in
an upcoming work \cite{upcoming}. 
At this point we just mention the important fact that both non-perturbative 
methods agree on the structure of the flow diagram, as it is depicted in Figure \ref{alphaflow}, 
and hence on the existence of the nontrivial fixed point and on the number of relevant 
directions at this fixed point. But they differ in the position of the fixed point, 
which has a positive $\tilde{\alpha}^*$ in case of the MCRG. This is in fact not surprising, 
since the position of the fixed point is not universal and because in the lattice 
calculations the higher order operators affect the flow, even though the applied RG procedure 
keeps track only of a truncated operator space, cf. \cite{upcoming} for more details.

Since $\alpha$ is not generated in this truncation, the system of two couplings 
effectively reduces to the one-parameter truncation that was investigated in
detail in \cite{Codello:2008qq}. While the critical 
value $\tilde{\zeta}^* = 16(N-2)/(45\pi^2)$ depends linearly on $N$, the critical 
exponent $\tfrac{d}{d\tilde{\zeta}}\beta_{\tilde{\zeta}}|_{\tilde{\zeta}^*}$ is independent of $N$:
it is $-16/15$ for all $N$. In this sense our computation is reminiscent of
the one-loop large-$N$ calculations \cite{Zinn}, apart from the small deviation 
of our critical exponent from the large-$N$ value $-1$. It is interesting to note 
that the running of $\zeta$ in the one-parameter truncation of the nonlinear model 
agrees exactly with the running that can be derived by means of the FRG if one regards 
the nonlinear model as the limit of a linear model with infinitely steep potential \cite{Codello:2008qq}.
However, while already a simple truncation of the linear model reproduces
reasonable results for the $N$-dependent critical exponents of the ${\rm O}(N)$-universality 
class, this $N$-dependence is apparently suppressed by taking the limit.

Let us briefly recall what is known about the critical properties of the nonlinear ${\rm O}(N)$ model in three dimensions:
The second order phase transition of the model is described by the nontrivial fixed point 
and its IR-relevant direction. The scaling of the critical coupling is directly related 
to the scaling of the correlation length: 
\begin{align}
 \nu = -\frac{1}{\Theta^*},
\end{align}
where $\Theta^*$ denotes the eigenvalue of the stability matrix evaluated at the fixed point, 
which corresponds to the relevant direction. The critical properties of linear and nonlinear ${\rm O}(N)$
models are intensively studied, see \cite{Zinn,Pelissetto,Ballesteros,Butera}, and it is generally 
believed that both theories belong to the same universality class. While there are 
Monte Carlo simulations of the nonlinear model which manifestly implement the nonlinearity 
of the target space and confirm this equivalence \cite{Ballesteros}, the analytic calculations 
rely either on an explicit  breaking of the symmetry and/or on an embedding of the 
target space in a linear space. An important motivation for the analytic approach 
presented here is to implement the isometries of the target space in every step 
of the calculation.

In order to become sensitive to the $N$-dependence of the critical exponent $\nu$,
we must include higher order operators in our manifestly covariant flow equation. 
This agrees with the finding of \cite{Codello:2008qq} that the derived $\beta$-functions 
of sigma models with different symmetric target spaces coincide in 
a simple truncation. They differ only if one takes higher order
operators into account, whose number and structure 
depends strongly on the specific type of model.

\begin{figure}
\includegraphics[width=0.48\textwidth]{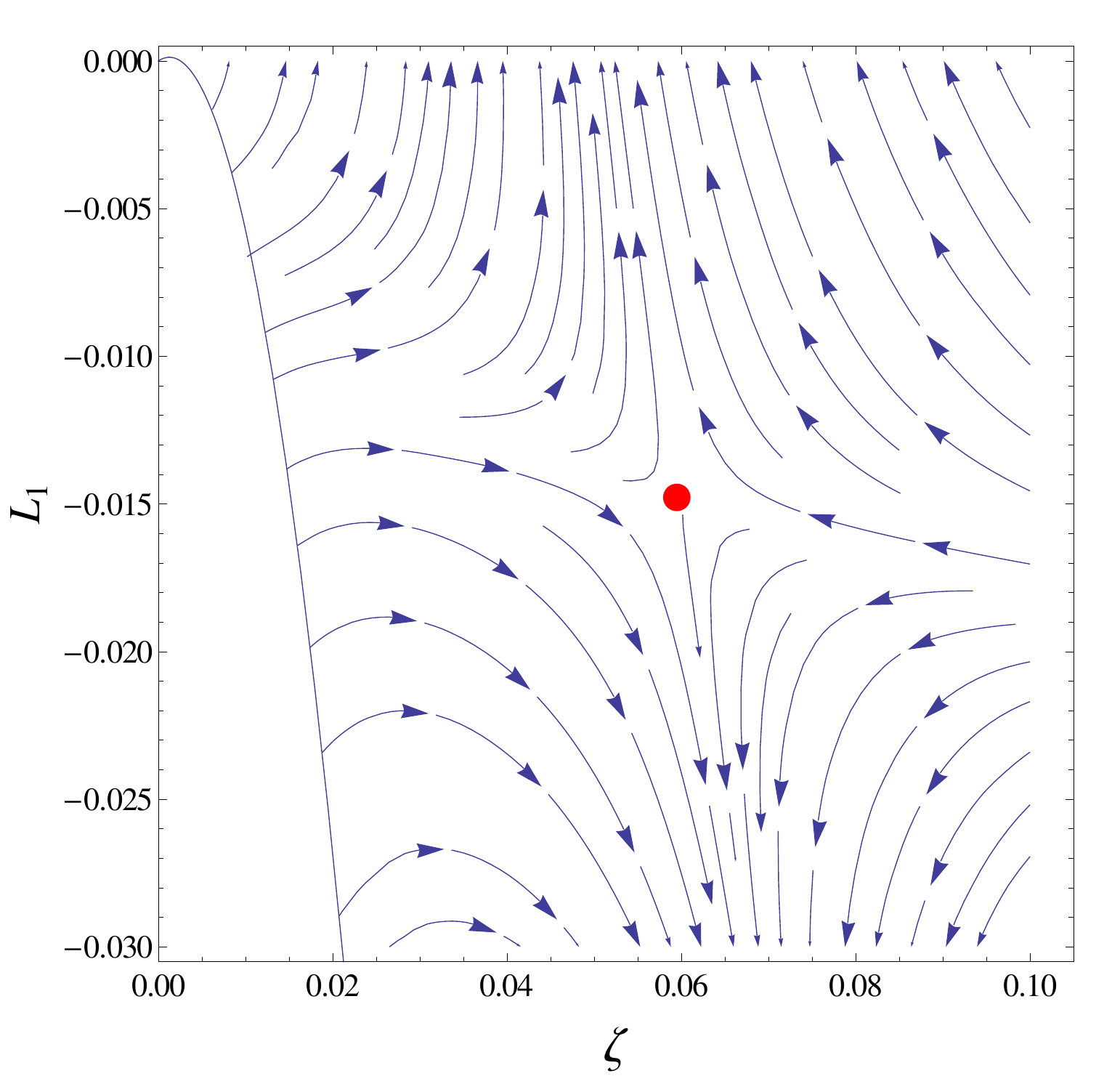}
\caption{The flow of the couplings $\tilde{\zeta}$ and $\tilde{L}_1$ towards the UV for $N=3$.
We have set $\tilde{\alpha}=\tilde{\alpha}^*=0$. The fixed point with one irrelevant direction 
is at $\tilde{\zeta}^*=0.059$ and $\tilde{L}_1^*=-0.013$.}
\label{Lflow}
\end{figure}

Hence we increase the truncation and include the operator $L_1 (h_{ab}\partial_{\mu} \phi^a \partial_{\nu} \phi^b)^2$. 
The resulting flow of the couplings $\tilde{\zeta}$ and $\tilde{L}_1$ of the ${\rm O}(3)$ model is depicted 
in Figure \ref{Lflow}, where we have set the irrelevant coupling $\tilde{\alpha}$ to $\tilde{\alpha}^*=0$.
It contains the nontrivial fixed point which was already discovered in the 
leading order truncation and which has only one relevant direction. 
The fixed point exists for all $N$, and while the critical 
value of $\tilde{L}_1$  is almost independent of $N$ and close to $-0.013$, 
the fixed point value $\tilde{\zeta}^*$ is an involved expression in $N$ which
for $N=3$ attains the values $0.059$. It increases  with $N$ such that it approaches a 
linear function with a slope of roughly $0.036$ for large $N$.
Similarly as for the leading order truncation the flow diagram has qualitatively the same 
structure as the diagram obtained by the MCRG \cite{upcoming}. The operator 
corresponding to $L_1$ is irrelevant in both approaches and its critical 
value $\tilde{L}_1^*$ is small in comparison to $\tilde{\zeta}^*$.

Actually there are additional fixed points in the truncation with coupling
$\zeta,\alpha$ and $L_1$, some 
with negative $\tilde{\zeta}^*$ and one with quite large values of $\tilde{L}_1^*$ and $\tilde{\zeta}^*$. 
These could be artifacts of our choice \eqref{optimized} of the cutoff
functions which may develop singularities for negative couplings.
We could not relate the additional fixed points to known critical properties 
of sigma models and their physical relevance remains unclear. We will therefore 
focus on the fixed point depicted in Figures \ref{alphaflow} and \ref{Lflow}.

\begin{figure}[h]
\includegraphics[width=0.45\textwidth]{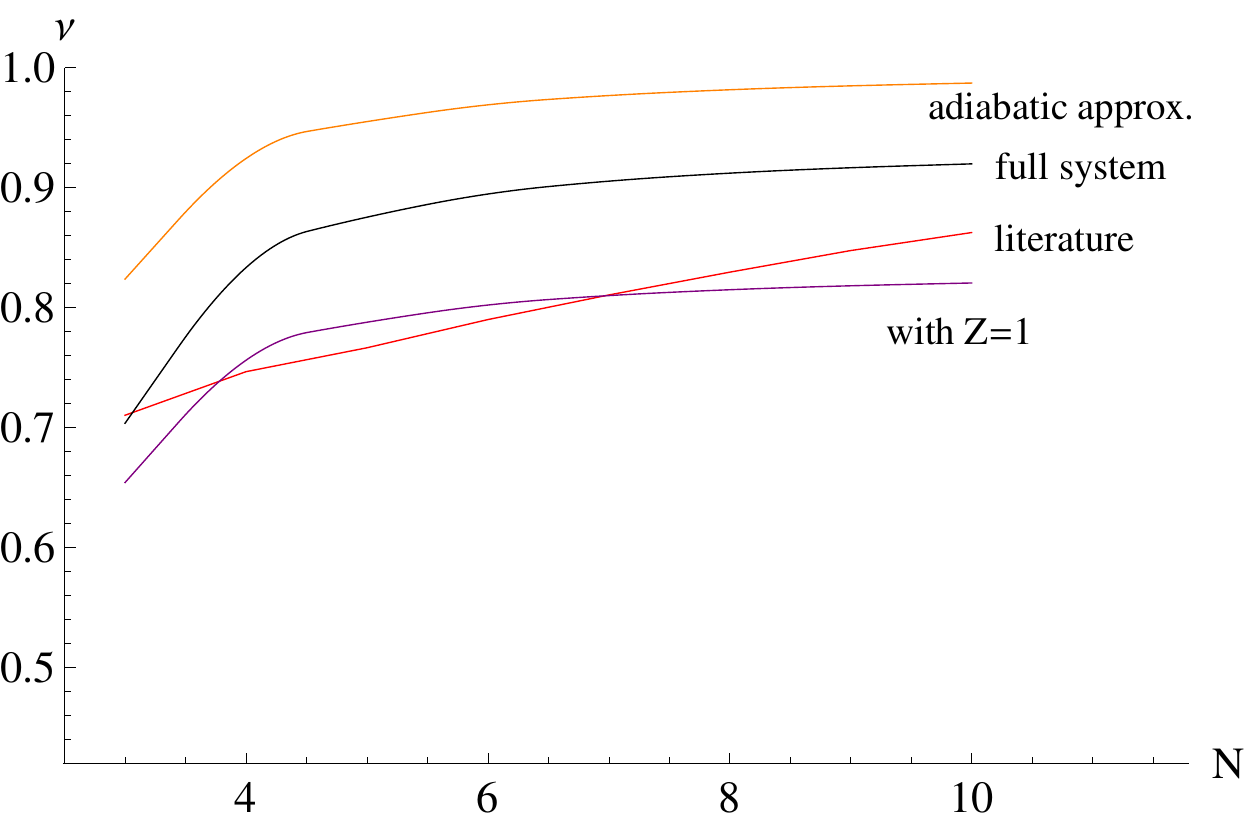}
\caption{The critical exponent $\nu$ as function of $N$, computed in the truncation $\{\zeta,\alpha,L_1\}$.
Depicted are the results of various approximations in comparison with average values from the literature.}
\label{nuComp}
\end{figure}

As anticipated, the inclusion of fourth order operators renders the exponent
$\nu$ sensitive to the dimension of the target space. The $N$-dependence of 
the exponent is depicted in Figure \ref{nuComp}, while the numerical values are given in Table \ref{nuTab}.
The values in the third row denoted by ``full system''  refer to calculations 
with the truncation $\{\zeta, \alpha, L_1\}$, in which the 
anomalous scaling $\eta$ of the fluctuation fields $\xi$ is taken into account. 
If one sets $Z\equiv 1$, one obtains the values in the second row of Table \ref{nuTab}.
If in addition one neglects the $k$-derivative of the couplings in $k\partial_k{\cal R}^k$
on the right hand side of the flow equation, that amounts to an
adiabatic approximation, then we obtain the values
in the first row of Table \ref{nuTab}.
At the fixed point the $k$-derivative of the couplings vanish such
that the approximation with $Z=1$ and the cruder adiabatic approximation
yield the same fixed point couplings.
\begin{table}[h!]
 \begin{tabular}{|c||c|c|c|c|c|c|}
\toprule
 $N$ & 3 & 4 & 6 & 8 & 10 & 20\\
\colrule
adiabatic approx. & 0.824 & 0.924 & 0.969 & 0.981 & 0.987 & 0.995\\
with $Z=1$ & 0.654 & 0.756 & 0.802 & 0.815 & 0.820 & 0.828\\
full system & 0.704 & 0.833 & 0.895 & 0.912 & 0.920 & 0.931\\
literature & 0.710 & 0.747 & 0.790 & 0.830 & 0.863 & 0.934\\
\botrule
 \end{tabular}
\caption{The critical exponent $\nu$ for various values of $N$ and
different approximations. The last row contains the best-known
values from the literature.}
\label{nuTab}
\end{table}
Since $\nu(N)$ is a rather involved and long expression, we tabulated only some
selected values in the rows of Table \ref{nuTab}.
For a comparison we added the last row which contains the values taken from the vast literature about the 
critical properties of the ${\rm O}(N)$ universality class. For $N=3$ and $N=4$ the values
are taken from the review \cite{Pelissetto}, which contains the results of many independent computations
of which we took the non-biased mean values. For $N>5$ we took the mean values of the
results in \cite{Butera,Kleinert,Antonenko}, which have been obtained by 
a high-temperature expansion, a strong-coupling expansion and six-loop RG expansion 
including a Pade-Borel resummation. The corresponding values deviate from each other 
by less than two percent.

The three truncations of the flow equation presented in this work yield a critical exponent $\nu$ 
whose $N$-dependence roughly agrees with the results in the literature. The values
obtained in the adiabatic approximation deviate considerably from the references 
values for small $N$, but show the correct large-$N$ asymptotic.
If one takes the running of the couplings in $k\partial_k{\cal R}^k$ into account, the results
for small $N$ improve significantly, especially if one neglects the wave function 
renormalization. In this case, however, $\nu(N)$ approaches for large $N$ 
the value  $5/6$ instead of the correct value $1$. If in addition one includes
the wave function renormalization, one obtains 
a critical exponent $\nu(N)$ which is closer to the reference value than 
the adiabatic result and whose asymptotic behavior is better behaved as in
the approximation with $Z=1$. The deviation from the best-known value is maximal
for $N=5$, where it is $14\%$,  and the asymptotic value is $15/16$ instead of $1$. 
This is in fact the value we found in the reduced truncation with just one coupling
and agreement originates from  $\lim_{N\to\infty} (\tilde{L}^*_1/\tilde{\zeta}^*)=0$.

It is not clear to what extent the inclusion of the wave function renormalization
improves the situation: on one hand a running $Z$ improves the asymptotic of $\nu(N)$ for
large $N$ and on the other hand $Z=1$ yields more accurate results for small $N$.
We included the wave function renormalization mainly for conceptual reasons
since the background and fluctuation fields are treated differently in the FRG 
formalism  and hence may possess different renormalization properties. This is
taken into account by admitting a running of $Z$. 

So far we did not consider the last contribution to the functional $\Gamma_k^b[\varphi,\xi]$
in \eqref{Eaction2} containing the mass parameter $m_k^2$. We included it in order 
to examine if terms that go beyond the ansatz of a ``single-field functional'' can 
improve the accuracy of the results. If we consider a truncation 
with couplings $\{\zeta,\alpha,L_1,m^2\}$ we find the same fixed 
point as before with slightly modified critical values and a positive mass parameter. 
However, the results for the critical exponent get worse rather than better and are 
close to the values of the adiabatic approximation. This is a
surprising finding and certainly requires a better understanding. For this purpose, the effects 
of higher-order terms in $\Gamma_k^b[\varphi,\xi]$ ought to be studied.

Let us finally include the remaining operator with four
derivatives $L_2(h_{ab}\partial_{\mu}\phi^a\partial^{\mu}\phi^b)^2$.
Although it is of the same order as the operators with couplings $L_1$ and
$\alpha$, it changes the flow such that there is no nontrivial fixed point for
the system with couplings $\{\zeta,\alpha,L_1,L_2\}$. This statement holds true 
for all possible modifications of the flow, i.e. in the adiabatic approximation, 
in the approximation with $Z=1$ and even if we include a mass parameter. In fact, already 
in the cruder truncation $\{\zeta,\alpha,L_2\}$ there is no nontrivial fixed point and 
it seems as if the renormalization of the coupling $L_2$ is not well-balanced.

One may wonder why the operator corresponding to $L_2$ destabilizes the renormalization
group flow. In the computation of the full effective action, the renormalization of an 
operator of a given order is always affected by operators of higher order. 
These contributions are lost in a truncation in which the higher order
operators are neglected.
In the present case the beta functions of $L_1$ and $L_2$ are quadratic functions, 
see \eqref{bL} in the appendix, and the coefficients of the polynomials must be fine-tuned 
such that both beta functions vanish.
We checked that the beta function of $L_2$, when evaluated at the 
fixed point for the subsystem consisting of all other couplings,
is nearly zero. Thus we expect that the inclusion of higher order terms 
will slightly modify the flow in a way that one recovers the fixed point and 
the information about the phase transition of the ${\rm O}(N)$ model, 
that we already detected in the truncation $\{\zeta,\alpha,L_1\}$.

However, there could be more subtle explanations why the flow of the operator
$(h_{ab}\partial_{\mu}\phi^a\partial^{\mu}\phi^b)^2$ does not lead to a stable fixed point.
We only want to mention two
possibilities: There were arguments brought forward recently \cite{bologna}
that the regularization procedure of the functional renormalization group
given by the introduction of $\Delta S_k$ may naturally require
a corresponding modification of the path integral measure, which in turn provides an additional 
term to the flow equation of the average effective action.
While this term yields only a renormalization of the vacuum energy in
theories with linearly realized symmetries,
it can affect the renormalization of nontrivial operators in theories whose path integral measure is field-dependent.

The second possibility is that, in order to find a stable fixed point for the full system,
one has to enlarge the truncation as dictated by hidden symmetries involving the background and the fluctuation fields.
To this day the background field method is the most effective way to deal with 
nontrivial field-space geometries in the framework of the FRG. Nevertheless, further 
studies maybe needed to understand better which truncations in terms of
background and fluctuation fields ought to be chosen in order to maintain the full 
reparametrization invariance of the theory. An ansatz that is based 
on the so-called Nielsen identities was presented recently in the context of gravity \cite{pawlowski} and it could 
be interesting to apply this approach to the nonlinear sigma model.

\paragraph*{A smooth $\alpha$-independent cutoff.}

The destabilization of the flow induced by the $L_2$-term does not seem to depend on a 
specific choice of regulator. For example, we used the alternative regulator $R_k[z]= k^{d+2}/z$ 
and confirmed the existence of nontrivial fixed points as well as the $N$-dependence 
of $\nu$ in the truncation $\{\zeta,\alpha,L_1\}$, but the fixed point still 
disappears if one includes $L_2$.
It was pointed out in \cite{Morris:2005ck} that there are concerns about
the application of an optimized cutoff like \eqref{optimized}
in the study of the linear sigma model beyond the local potential approximation.
This motivates us to briefly discuss our system using a smooth cutoff like the exponential one
\begin{equation}
R_k[z]= A Z_k \frac{z}{{\rm e}^{z/k^{2}}-1}\,,
\end{equation}
where $A$ is an external parameter that can be tuned
following the so-called ``principle of minimum sensitivity'' \cite{Canet:2003qd}.
For a generic choice of $A$ it is possible to evaluate numerically
the $Q$-functionals \eqref{Qfunctionals} and therefore determine the beta functions of the theory with arbitrary precision.
Using the exponential cutoff, we studied the complete system of beta functions at $N=3,4,5,10,100,1000$, including $L_2$, in the parameter range $A\in (0,10]$, but could not stabilize the nontrivial fixed point.
Resorting to the approximation $L_2=0$, a nontrivial FP with one attractive direction appears as a function of the two parameters.
The numerical results for the scaling exponent $\nu$ of the correlation length
are in qualitative agreement with the optimized cutoff results for every value bigger than $A\simeq 0.18$,
that we analyzed in the parameter space with accuracy $\Delta A=0.01$.
Since $\nu$ is computed as a numerical function of $A$ in the form $\nu(A)$,
one could apply the principle of minimum sensitivity, i.e. finding the best value for $A$ as a local minimum $A^\star$ of $\nu(A)$.
However, the result seems to indicate that such a minimum does not exists, since the function $\nu(A)$
appears to be (slowly) monotonically increasing to our accuracy.

\section{Conclusion}\label{conclusion}

In this article we applied the functional renormalization technique based
on a scale dependent effective action to investigate the renormalization group 
flow of the nonlinear ${\rm O}(N)$-model.
The flow is formulated in a manifestly reparametrization invariant way,
so that the results do neither depend on any specific choice of coordinates on
the target sphere, nor on an implicit embedding of the nonlinear model into the linear one.
Since the symmetries of the theory are realized nonlinearly,
we adopted a geometric formulation where a background (base-point) dependence 
is introduced in order to maintain the covariance of the model.
The background field is used to construct a quadratic infrared cutoff term for the fluctuations,
whose purpose is to allow us to effectively integrate out the ultraviolet modes
while simultaneously respecting the symmetries of the model.
The main achievement is the construction of a scale-dependent effective action,
that is ${\rm O}(N)$ invariant for both the transformations of the background and the quantum field.
The consistency of our formalism was underlined by the appearance of nontrivial relations between 
the renormalization flow of background and fluctuation operators.

The model has been studied using a truncation ansatz
that includes all possible covariant operators up to fourth order in the derivatives.
The beta functions of the theory are identified with the derivatives of the couplings 
w.r.t. the logarithm of the RG scale. They provide important informations 
about the phase diagram of the model and we concentrated our discussion
in particular on the fixed points of the renormalization flow, because they are 
associated to second order phase transitions. The scaling of the correlation length is 
a universal property and is directly related to the critical exponent corresponding to the
relevant direction. Our explicit calculations focused on three dimensions,
in which the critical exponents of the model are well-studied and provide a rich literature 
that can serve as a reference for comparison.

Our investigation did not immediately point to the existence of such universal behavior,
because we did not find a suitable fixed point for the full fourth order system.
However, we discovered that in the restricted subspace of couplings,
where one of the higher derivative couplings is set to zero ($L_{2}=0$), a fixed point for all $N$ emerges. 
It exhibits one relevant direction, which is already present in an one-parameter 
truncation, and it is such that the inclusion of further couplings ($L_1$ and $\alpha$) only adds irrelevant directions.
The critical value 
of the coupling $\alpha$ is identically zero in a fourth-order expansion. However, the one-parameter 
truncation is not sensitive to the $N$-dependence of the critical exponent $\nu$, which requires higher-order
operators. The results we obtain for $\nu$ in the truncation $\{\zeta,\alpha,L_1\}$ agree qualitatively
with the pre-existing literature, but show some numerical difference that is likely due to the 
limited truncation ansatz.
We tested the presence of this fixed point against various approximations
and choices of the coarse-graining scheme, to find that it is a very stable result.

The renormalization properties of the nonlinear ${\rm O}(N)$ model have also been studied
by means of the Monte Carlo Renormalization Group and will be presented in detail elsewhere \cite{upcoming},
where a discretized version of the action \eqref{Eaction1} is considered as lattice action.
The quantitative comparison of the results of FRG and MCRG is a delicate topic
that has to be addressed with care. In fact, one must bear in mind that our truncation ansatz
\eqref{Eaction1} is \emph{not} a bare action \cite{Ellwanger:1997tp}.
The nature of the ``mass'' regularization of FRG, as opposed to that of the lattice which is mass-independent,
is expected to produce a nontrivial relation between the renormalized couplings of the two methods \cite{Manrique:2008zw}.
However, we do expect the phase diagrams of FRG and MCRG to share the same qualitative properties
(namely the existence of a fixed point with one UV-attractive direction).
The MCRG flows that are computed for the fourth-order expansion of the model are (apart from the coupling $L_2$)
in good qualitative agreement with the results presented here \cite{upcoming}.

As mentioned, a nontrivial fixed point appears in our calculations only if one 
coupling, $L_{2}$, is neglected. We strongly believe that this is not a pathology 
of the computation, but rather it is due to the limited truncation considered and the quadratic 
structure of the computed beta functions.
However, we also admit the possibility that the operator corresponding to the coupling $L_{2}$ may have a special
role in the nature of this phase transition. Therefore further investigations are required in order to obtain a deeper understanding.

\section{Acknowledgments}\label{acknowledgments}

We would like to thank  Holger Gies, Jan Pawlowski, Martin Reuter, Bj{\"o}rn Wellegehausen and
Daniel K{\"o}rner for useful discussions and
the latter two for the collaboration on an accompanying numerical
project on the renormalization group flow of sigma models. RF and OZ benefited from
discussions with Alessandro Codello, Maximilian Demmel and Roberto Percacci.
This work has been supported by the DFG Research Training Group ``Quantum
and Gravitational Fields'' GRK 1523.
The research of OZ is supported by the DFG
within the Emmy-Noether program (Grant SA/1975 1-1).

\appendix

\section{Beta functions of $L_1$ and $L_2$}
In this appendix we give the explicit form of the beta functions of the couplings $L_1$ and $L_2$.
These have been omitted from the main text because of their size.
Together with \eqref{beta functions} they complete the system of beta functions of the couplings of \eqref{Eaction1}.
We computed:
\begin{widetext}
\begin{align}
\label{bL}
 \beta_{L_1}
 =& [ (2N-5) L_1 + 2 L_2 - \alpha] Q_{\frac{d}{2},2}
 + \Bigl[(2(N+4) + 4 d + d^2) L_1^2 + 8 L_2^2 + 4 (d+2) L_2 \alpha+ d(d+2) \alpha^2
  \nonumber\\
  &
 + 2 L_1 (2 (d+6) L_2 + (d^2 + 3 d +2) \alpha)\Bigr] Q_{\frac{d}{2}+2,3}
 + 2 [(d+1)L_1 + 2 L_2 + d \alpha]\zeta Q_{\frac{d}{2}+1,3} + \zeta^2 Q_{\frac{d}{2},3} + \frac{1}{6} Q_{\frac{d}{2}-2,1} \,, 
 \nonumber
 \\
 \beta_{L_2}
 = 
 &
 \Bigl[ (d^2 (N-1) + 2 d (N+1)+12)L_2^2
 + (N + 2 d + 6) L_1^2 - 2 (d+2)(2 + d (N-2)) L_2 \alpha
 \nonumber\\ 
 &
 + 2 (d^2 + 2 d + 4 + N(d+2))L_1 L_2
 - 2(d+2)(N-1 + d) \alpha L_1
 + d (d+2) (N-3) \alpha^2\Bigr] Q_{\frac{d}{2}+2,3}
 \nonumber\\
 &
 -2 [L_2 ((N-2)d+2) + L_1 (N-1 + d)
 - (N-3)d\alpha] \zeta Q_{\frac{d}{2}+1,3}
 + [L_1 +2(N-3) L_2- (N-3) \alpha] Q_{\frac{d}{2},2}
 \nonumber\\
 &
 +(N-3) \zeta^2 Q_{\frac{d}{2},3}
 -\tfrac{1}{6} Q_{\frac{d}{2}-2,1}
 \,.
\end{align}
\end{widetext}


\end{document}